\documentclass[12pt]{article}
\usepackage{epsfig,amssymb,amsmath,psfrag,subfigure,rotate,color,wasysym}

\allowdisplaybreaks

\def\XXint#1#2#3{{\setbox0=\hbox{$#1{#2#3}{\int}$ }
\vcenter{\hbox{$#2#3$ }}\kern-.6\wd0}}

\def \be  {\begin{equation}}
\def \ee  {\end{equation}}
\def \ba  {\begin{eqnarray}}
\def \ea  {\end{eqnarray}}
\def \baa {\begin{eqnarray*}}
\def \eaa {\end{eqnarray*}}
\def \lab #1 {\label{#1}}

\newcommand\re[1]{(\ref{#1})}
\def\d{\hbox{{d}\kern-.20em\hbox{l}}}

\def \matrx #1 {\left(\begin{array}{cc} #1 \end{array}\right)}

\newcommand{\bit}[1]{\mbox{\boldmath$#1$}}

\begin{document}

\begin{titlepage}

\thispagestyle{empty}

\vspace*{3cm}

\centerline{\large \bf Balancing act: multivariate rational reconstruction for IBP}
\vspace*{1cm}

\centerline{\sc A.V. Belitsky$^a$, A.V.~Smirnov$^{b,c}$, R.V.~Yakovlev$^{b,c}$}

\vspace{10mm}

\centerline{\it $^a$Department of Physics, Arizona State University}
\centerline{\it Tempe, AZ 85287-1504, USA}

\vspace{5mm}

\centerline{\it $^b$Research Computing Center, Moscow State University}
\centerline{\it 119992 Moscow, Russia}

\vspace{5mm}

\centerline{\it $^c$Moscow Center for Fundamental and Applied Mathematics}
\centerline{\it 119992 Moscow, Russia}

\vspace{20mm}

\centerline{\bf Abstract}

\vspace{5mm}

We address the problem of unambiguous reconstruction of rational functions of many variables. This is particularly
relevant for recovery of exact expansion coefficients in integration-by-parts identites (IBPs) based on modular 
arithmetic. These IBPs are indispensable in modern approaches to evaluation of multiloop Feynman integrals by means
of differential equations. Modular arithmetic is far more superior to algebraic implementations when one deals with 
high-multiplicity situations involving a large number of Lorentz invariants. We introduce a new method based on balanced 
relations which allows one to achieve the goal of a robust functional restoration with minimal data input. The technique is 
implemented as a Mathematica package {\tt Reconstruction.m} in the FIRE6 environment and thus successfully demonstrates 
a proof of concept.

\end{titlepage}

\setcounter{footnote} 0

\renewcommand{\thefootnote}{\arabic{footnote}}

\section{Introduction}

Integration-by-parts identities (IBPs) \cite{Chetyrkin:1981qh}, see also Chapter 6 of the book \cite{Smirnov:2012gma}, are an indispensable tool to 
reduce an arbitrarily large set of Feynman integrals to a finite set \cite{Smirnov:2010hn} of the so-called Master Integrals (MIs). With development of 
powerful computers, a number of programs had become available over the past twenty years to handle algebraically a plethora of IBPs. The bulk of 
them \cite{Anastasiou:2004vj,Smirnov:2019qkx,Studerus:2009ye,Maierhofer:2017gsa,Klappert:2020nbg} (for the exception of \cite{Lee:2012cn}) are 
based on the renowned Laporta algorithm \cite{Laporta:2000dsw} of Gauss elimination for a given choice of priority ordering.

IBPs form a system of linear algebraic equations with matrix coefficients whose elements are generically given by rational functions
\begin{align}
\label{Fblackbox}
F(\bit{x}) = \frac{P (\bit{x})}{Q (\bit{x})}
\, .
\end{align}
Here $P$ and $Q$ are polynomials of (typically) different degrees in the number of space-time dimension $d$ and $(L-1)$ Lorentz invariants. These
are cumulatively denoted by the vector of variables $\bit{x} = (x_1, \dots, x_L)$. 

The main problem with the Laporta reduction is the {\sl swelling of intermediate expressions} when one performs the elimination in an algebraic manner 
and which, in turn, leads to severe computer performance issues, as these become incredibly time consuming to manipulate with and hard to store in a 
memory. The final form of the coefficients is, however, rather concise in length (several orders of magnitude less than their intermediate form). 

A natural way out would be to perform all calculations numerically and then attempt their analytic reconstruction. The use of floating point arithmetic is a 
no-starter in this context however since one is aiming at an exact reconstruction of the expansion coefficients. The use of rational numbers instead are 
plagued by problems akin to ones emerging in analytic treatments since they require arbitrary precision arithmetic which is time consuming. So an 
idea was put forward in Ref.\ \cite{vonManteuffel:2014ixa} to use numerical techniques over finite fields in computer algebra manipulations of IBPs. 
These are obviously advantageous compared to the ones we just alluded to above since, as the name suggests, there is only a finite number of elements 
involved (with well-defined inverses) and they can be represented by machine-size integers. Their disadvantage, however, is information loss along the 
way and thus the necessity to use several finite fields for back recovery of rational numbers. However, these efforts are far less demanding than the 
direct use of the latter. This rational reconstruction from its images in several fields has been known for quite a while and is implemented through the 
Extended Euclidean Algorithm \cite{Wang1981,Monagan2004} which relies on the Chinese Remainder Theorem \cite{Gathen2013}.

Having addressed the proper numbers for numerical calculations, the main problem then consists in the actual reconstruction of {\sl the black box} 
\re{Fblackbox} from sample data (aka black box points) with high probability of success. While univariate methods date back to more than a century 
ago as celebrated Newton and Thiele interpolations \cite{AbrSte64} for polynomial and rational functions, respectively, multivariate techniques are 
relatively new. Interpolations for sparse\footnote{It refers to a given upper bound on the number of terms in $P$ and $Q$ of Eq.\ \re{Fblackbox}.} 
functions of many variables were addressed in Refs.\ \cite{BenOrTiw88,Zippel90,Grigoriev1994,Kaltofen88,Kleine2005,HomoMulti2011a,HomoMulti2011b,HuaGao17}. 
Reconstruction methods for dense\footnote{I.e., unconstrained number of terms in $P$ and $Q$.} multivariate functions are more rare. A generalization 
of the Thiele method was proposed and implemented with the release of the FIRE6 IBP reduction framework in Ref.\ \cite{Smirnov:2019qkx}. However, 
it can hardly be used beyond two variables, since it faces severe computational challenges. The most prominent sparse reconstruction is based on 
the so-called homogeneous interpolation \cite{HomoMulti2011a,HomoMulti2011b}. It was adopted and remastered in Refs.\ \cite{Peraro:2016wsq} and 
\cite{Klappert:2019emp} for scattering amplitude problems through the {\tt FiniteFlow} and {\tt FireFly} packages, respectively. {\tt FireFly} was also 
integrated with Kira2.0 IBP program in Ref.\ \cite{Klappert:2020nbg}. 

In circumstances when the time of numerical computations of a single sample point is comparable to the one of the total reconstruction, it is very important
to have a method which requires minimal initial data set. In this regards, the homogeneous multivariate reconstruction is indeed a viable tool, however, it 
requires sufficiently high number of probes for successful sought-after reconstruction as it is very sensitive to the total power of polynomials building up
the numerator/denominator in Eq.\ \re{Fblackbox}. This technique will be used by us as a benchmark for the method which we propose in this 
work\footnote{A preliminary version of the method, which we introduce and explore in the current paper, was discussed in Ref.\ \cite{Chuh20}.}. We will 
demonstrate that it will allow us for a more economical computational efforts compared to the former since it requires less black box probes.

Our subsequent presentation is organized as follows. In the next section, we recall classical methods of univariate interpolations. Then in 
Sect.~\ref{BalancedReconSection}, we present our new framework based on the so-called {\sl balanced} reconstruction. We address the
issue of the most optimal ordering in Sect.~\ref{OptimalSection} and then, in Sect.~\ref{BenchmarkSection}, compare our method with the
homogenous one. In Sect.~\ref{ReconFIRESection}, we introduce a Mathematica code integrated with the FIRE6 program for the balanced
reconstruction of IBPs and give a thorough example in Sect.~\ref{ExampleSection}. Finally, we conclude and discuss future directions. This 
manuscript is also accompanied by the Mathematica script code {\tt Reconstruction.m} and the notebook {\tt doublebox.nb} detailing all recovery 
steps, as well as syntax and commands, for a typical process of IBP reduction.

\section{Classical univariate reconstructions}

To start with, let us recall two classical results used in polynomial and rational interpolation of functions of a single variable $x$, which we build upon
in the following sections. These are known as Newton and Thiele methods, respectively. 

Throughout this paper, we will be adhering to the following notations: $N$ will be the number of samples for the Newton method, while $T$ will be 
the number of data points for the Thiele method. Greek letters will denote integer labels of fixed numerical values of variables, e.g., $x_\alpha$.

\subsection{Newton method}

The most basic method of polynomial interpolation of a function $f(x)$ is based on the so-called Newton interpolating polynomials $f_N (x)$ on $N$ 
distinct sampling data points $x_\alpha$ with $\alpha = 1, \dots, N$,
\begin{align}
f_N (x) 
&= {\rm Newton}_x [f (x), N]
\\
&
\equiv 
a_1 + (x - x_1) \Big[ a_2 + (x - x_2) \big[ a_3 + (x - x_3) \left[ a_4 + \dots \right] \big] \Big]
\, . \nonumber
\end{align}
The accompanying coefficients are defined recursively through the divided differences
\begin{align}
a_1 
&= f (x_1) \, ,\nonumber\\
a_2
& 
=
[f(x_1) , f (x_2)]
\equiv
\frac{f(x_1) - f (x_2)}{x_1 - x_2} \, ,\nonumber\\
a_3
& 
=
[f(x_1) , f(x_2),  f (x_2)]
\equiv
\frac{[f(x_1) , f (x_2)] - [f(x_2) , f (x_3)]}{x_1 - x_3} \, ,\nonumber\\
a_4
& 
=
[f(x_1) , f(x_2),  f(x_3),  f (x_4)]
\equiv
\frac{[f(x_1) , f (x_2) , f(x_3)] - [f(x_2) , f (x_3), f(x_4)]}{x_1 - x_4} \, ,\nonumber\\
&
\vdots \nonumber\\
a_N
& 
=
[f(x_1) , f(x_2),  f(x_3), \dots, f(x_N)]
\nonumber\\
&
\equiv
\frac{[f(x_1) , f (x_2) , f(x_3), \dots, f(x_{N-1})] - [f(x_2) , f (x_3) , f(x_4), \dots, f (x_N)]}{x_1 - x_N} \, .
\end{align}

Obviously, if the function of interest $f (x)$ is known to be a polynomial of a predetermied degree ${\rm deg} [f (x)]$ to start with, one can 
unambiguously reconstruct it by sampling in
\begin{align}
N = {\rm deg}_x [f(x)] + 2
\end{align}
points. The last one being the control probe such that the function does not change by adding more data points
\begin{align}
f_{N + 1} (x) = f_N (x) = f (x)
\, .
\end{align}
The advantage of the Newton method compared to the naive power expansion with unknown coefficients is the fact that an addition on a
new data point does not necessitates reevaluation of all of the coefficients from scratch.

\subsection{Thiele method}

Rational interpolation of a function $f(x)$, which typically yeilds a better approximation than the above polynomial interpolation, is achieved 
with the help of the Thiele continued fraction on $T$ black box probes $x_\alpha$ with $\alpha = {1, \dots, T}$
\begin{align}
f_T (x) 
&= {\rm Thiele}_x [f (x), T]
\\
& 
\equiv
b_0 + (x - x_1) \left[ b_1 + (x - x_2) \left[ b_2 + (x - x_3)\left[ b_4 + \dots \right]^{-1} \right]^{-1}\right]^{-1}
\, , \nonumber
\end{align}
and the coefficients being determined by the following relations
\begin{align}
b_1 
&= f (x_1) \, ,\nonumber\\
b_2
& 
=
[f(x_1) , f (x_2)]_r
\equiv
\frac{x_1 - x_2}{f(x_1) - f (x_2)} \, ,\nonumber\\
b_3
& 
=
[f(x_1) , f(x_2),  f (x_2)]_r
\equiv
\frac{x_1 - x_3}{[f(x_1) , f (x_2)]_r - [f(x_2) , f (x_3)]_r} \, ,\nonumber\\
b_4
& 
=
[f(x_1) , f(x_2),  f(x_3),  f (x_4)]_r
\equiv
\frac{x_1 - x_4}{[f(x_1) , f (x_2) , f(x_3)]_r - [f(x_2) , f (x_3), f(x_4)]_r} \, ,\nonumber\\
&
\vdots \nonumber\\
b_N
& 
=
[f(x_1) , f(x_2),  f(x_3), \dots, f(x_N)]_r
\nonumber\\
&
\equiv
\frac{x_1 - x_N}{[f(x_1) , f (x_2) , f(x_3), \dots, f(x_{N-1})]_r - [f(x_2) , f (x_3) , f(x_4), \dots, f (x_N)]_r} \, ,
\end{align}
closely related to the reciprocal differences. 

Again, if a function $f(x)$ is known to be rational to begin with, the method allows to exactly reconstruct it by sampling in 
$T$ points, with the latter determined by the following estimate (including the control probe)
\begin{align}
\label{ThieleNumber}
T \simeq 2 \times \mbox{max} \left\{ \mbox{deg}_x [\mbox{Numerator} [f(x)]] , \mbox{deg}_x [\mbox{Denominator} [f(x)]] \right\} + 1
\, .
\end{align}
Akin to the Newton method, Thiele reconstruction does not require recalculation of all the $b_j$'s with every new sample added, contrary to
other methods, for instance, the so-called barycentric interpolation \cite{Barycentric2007}. Sometimes, the above algorithm may yield a 
vanishing denominator, for instance, when two successive points possess the same dependent value or when one samples three 
collinear successive data points. In these circumstances, all one has to do is to perturb data points ever so slightly to get rid of the problem.

\section{Balanced reconstruction}
\label{BalancedReconSection}

With the above lightning overview of univariate interpolations behind us, let us introduce a new approach to multivariate rational 
reconstruction, which we dub {\sl the balanced reconstruction}.

Consider a rational multivariate function of $L$ variables
\begin{align}
F =  F (\bit{x})
\, , \quad\mbox{with}\quad \bit{x} = (x_1, \dots, x_L)
\, .
\end{align}
Let us spit the total vector of variables $\bit{x}$ into three orthogonal vector subspaces
\begin{align}
\bit{x} = (\bit{d}, x_j, \bit{r})
\, , \quad\mbox{with}\quad \bit{d} = (x_1, \dots, x_{j-1}) \, , \quad \bit{r} = (x_{j+1}, \dots, x_L)
\, .
\end{align}
with the $\bit{d}$-vector taking on the meaning of analytically reconstructed, or (d)one, variables, $x_j$ being the variables 
under consideration and $\bit{r}$ being the (r)emainder. We designate the function $F (\bit{x})$ with $\bit{d}$ reconstructed 
variables as
\begin{align}
F_{\bit{\scriptscriptstyle d}} (\bit{d}, x_j, \bit{r})
\, .
\end{align}

Before one starts the recovery algorithm, one has to get an estimate on the minimal number of sampling points needed for
successful reconstruction. This is accomplished by performing the univariate Thiele restoration for each variables $x_j$ from 
the vector $\bit{x}$ with all others $\bit{x}\backslash x_j$ kept fixed, yielding a value $T_j$ for a stable reconstruction. These 
are then used to get the minimal number of sample data points needed
\begin{align}
\{ x_{1, \alpha} , \alpha = 1, \dots, T_1 ; \quad x_{2, \beta}, \beta = 1, \dots, N_2 ; \quad \dots \quad; \quad x_{L, \gamma}, \gamma = 1, \dots, N_L \}
\, , 
\end{align}
with 
\begin{align}
\label{BalancedNewtonNumber}
N_j \simeq [T_j/2]
\end{align}
to be explained below (see Sect.~\ref{OptimalSection}).

The algorithm consists in the following steps.
\begin{enumerate}
\item \label{FirstStep} Numerically compute values of the function $F$ with fixed values of all variables in their respective ranges, 
determined from the preliminary estimates alluded to above,
\begin{align}
x_{1, \alpha} \, ,\quad \alpha = \{1, \dots, T_1 \}
\, , \qquad
(x_2, \dots, x_L)_\beta  \, ,\quad \beta = \{1,\dots, N_{2, \dots, L} \}
\, .
\end{align}
\item The first variable $x_1$ is reconstructed by means of the Thiele method,
\begin{align}
F_{x_1} (x_1, \bit{r}_\beta) = {\rm Thiele}_{x_1} [F (x_1, \bit{r}_\beta), T_1]
\, .
\end{align}
\item \label{NewtonTables} Let $\bit{d}$ be the vector of already reconstructed variables [i.e.,  $\bit{d} = (x_1)$ after the first step]. 
Collect tables of the function 
\begin{align}
F_{\bit{\scriptscriptstyle d}} (\bit{d}, x_{j, \alpha}, \bit{r}_\beta)
\end{align}
with $\alpha \in \{1, \dots, N_j\}$ and $\beta \in \{1,\dots, N_r\}$ from the above two steps. All other variables $\bit{x}\backslash x_1$ 
are handled by the {\sl balanced} Newton method as follows. 
\item \label{BalancingTables} Compute values of the function $F = F_{\bit{\scriptscriptstyle d}} (\bit{d}_0, x_{j,\alpha}, \bit{r}_\beta)$ for a single fixed 
numerical value of the vector of already done variables $\bit{d} = \bit{d}_0$, $T_j$ values of the variable $x_j$ under reconstruction $x_{j,\alpha}$, 
$\alpha \in \{1, \dots, T_j\}$, and $N_r$ values of the rest $\bit{r}_\beta$, $\beta \in \{1,\dots, N_r\}$ (same as in step \ref{FirstStep}). Thiele reconstruct 
$x_j$ from the set $F_{\bit{\scriptscriptstyle d}}  (\bit{d}_0, x_{j,\alpha}, \bit{r}_\beta)$ obtaining
\begin{align}
\mbox{\sl the \ balancing \ tables:} 
\qquad 
F_{\bit{\scriptscriptstyle d}, x_j}  (\bit{d}_0, x_j, \bit{r}_\beta) = F_{\bit{\scriptscriptstyle d}} (\bit{d}_0, x_j, \bit{r}_\beta)
\end{align}
for the variable $x_j$. Notice that this reconstruction step is univariate in the variable $x_j$.
\item Balance the set of values $F_{\bit{\scriptscriptstyle d}}  (\bit{d}, x_{j, \alpha}, \bit{r}_\beta)$ computed in step 
\ref{NewtonTables} with the balancing tables from step \ref{BalancingTables} by evaluating
\begin{align} 
V (\bit{d}, x_j, \bit{r}_\beta) 
=
\frac{
F_{\bit{\scriptscriptstyle d}}  (\bit{d}, x_{j, \alpha}, \bit{r}_\beta) 
\times
F_{\bit{\scriptscriptstyle d}, x_j}  (\bit{d}_0, x_j, \bit{r}_\beta) 
}{
F_{\bit{\scriptscriptstyle d}}  (\bit{d}_0, x_{j, \alpha}, \bit{r}_\beta) 
}
\, .
\end{align}
\item Factorize $V (\bit{d}, x_j, \bit{r}_\beta)$  into the numerator and denominator and separately Newton-re\-cons\-truct them individually in $x_j$ 
from the set of sample points $x_{j, \alpha}$ with $\alpha \in \{1, \dots, N_j\}$,
\begin{align}
\label{BalancedNewtonRecon}
F_{\bit{\scriptscriptstyle d}, x_j}  (\bit{d}, x_j, \bit{r}_\beta) 
=
\frac{
{\rm Newton}_{x_j} [ \mbox{Numerator} [V (\bit{d}, x_{j}, \bit{r}_\beta) ] , N_j]
}{
{\rm Newton}_{x_j} [ \mbox{Denominator} [V (\bit{d}, x_{j}, \bit{r}_\beta) ], N_j]
}
\end{align}
\item Proceed to step \ref{NewtonTables} for the next variable $x_{j+1}$. If $j = L$, the reconstruction stops.
\end{enumerate}
Having introduced the algorithm, let us proceed with its optimization.

\section{Optimal ordering}
\label{OptimalSection}

The order of the balanced reconstruction for multivariate functions is crucial for the overall execution speed of the algorithm 
as it affects the number of required black box probes $P$ for a robust recovery. Thus, the minimization of the value of $P$ serves 
as the main criterium for optimization. As a rough, naive estimate, it suffices to use the following rule of thumb: start with the variables 
requiring the lowest number of Thiele samples $T_j = {\rm min} \{ T_k, k = 1,\dots, L\}$ and then proceed in the order of their growth 
(though, their order is not very relevant)
\begin{align}
\label{RoughNumber}
P_{\rm naive} \simeq T_j \prod_{k\neq j}^L [T_k/2]
\, .
\end{align}
However, a more accurate value of $P$ was found experimentally by minimizing the product
\begin{align}
\label{MinPsample}
P_{\rm balance} = \mbox{min} \left\{ N_L \Bigg[ N_{L-1} \Big[ \dots \big[ N_2 (N_1 + D_1) + D_2 \big]\dots \Big] + D_{L-1} \Bigg]+ D_L \right\}
\, ,
\end{align}
where 
\begin{align}
D_j = \mbox{max}(T_j - N_j, 0)
\, ,
\end{align}
expressed in terms of the minimal number of the Thiele  \re{ThieleNumber} and balanced Newton \re{BalancedNewtonNumber} sampling 
points, respectively. While the value of $T$ is self-explanatory, $N$ is determined from
\begin{align}
\label{BalancedNewtonNumerEstimate}
N \simeq \mbox{max} \left\{ \mbox{deg}_x [\mbox{Numerator} [f(x)]] , \mbox{deg}_x [\mbox{Denominator} [f(x)]] \right\} + 2
\, ,
\end{align}
as it is clear from Eq.\ \re{BalancedNewtonRecon} with the addition of $2$ needed for restoration of an overall constant and control probe for 
correctness of recovery. A rigorous proof of the above estimate \re{MinPsample} would be very welcome.

The number of different combinations one has to compare in order to determine\ \re{MinPsample} is set by the number $L$ of components 
in $\bit{x}$ and equals to the number of inequivalent permutations $L!$. Even for $L = 10$, it is the minuscule $3,628,800$ compared to the
staggering, e.g., 11 trillion operations per second that, for instance, an M1 Max silicone can perform.

\section{Benchmark and comparison}
\label{BenchmarkSection}

As a benchmark, let us confront our new approach with the method of multivariate homogeneous interpolation 
\cite{HomoMulti2011a,HomoMulti2011b} with its reincarnation relevant to the dense rational reconstruction in 
Ref.\ \cite{Peraro:2016wsq}. The main idea consists in the homogeneous rescaling of all components of the 
vector $\bit{x}$ and introduction of a new function $h(z, \bit{x})$
\begin{align}
\bit{x} \to z \bit{x} \, , \qquad h(z, \bit{x}) = F(z \bit{x})
\, ,
\end{align}
such that one can clearly separate its numerator $P$ from the denominator $Q$. The algorithm then consists in just three steps.
\begin{enumerate}
\item Thiele reconstruct the variable $z$ with $T_z$ black box probes \re{ThieleNumber} for arbitrary fixed values of all other variables $\bit{x}_0$.
\item \label{Newton} Separate its numerator $P (z \bit{x}_0)$ and the denominator $Q (z \bit{x}_0)$ and Newton reconstruct them separately in $x_j$ 
variable on $N_j$ sample points given in Eq.\ \re{BalancedNewtonNumerEstimate}.
\item Proceed to step \ref{Newton} for the next variable $x_{j+1}$. If $j = L-1$, the algorithm stops.
\end{enumerate}
A few of comments are in order. First, the method requires generalization of the original function by introducing a new variable. Then, however, 
there is no need to reconstruct the last variable $x_L$ since it can be recovered using homogeneity. Second, there is an unpleasant subtlety in its 
application to denominators not possessing a constant term, which then vanishes for the point $\bit{x} = 0$, and the rational function becomes 
singular. If this is the case, one has to perform ad hoc shift of all variables \cite{HomoMulti2011a,HomoMulti2011b} and only then apply the above 
algorithm. This could potentially result in a more elaborate reconstruction process though. Third, the advantage of this method is the complete 
democracy among different ordering of variable reconstructions, there is not a preferred one as compared to our balanced algorithm that we 
advocated for above. This immediately provides an estimate on the number of black box probes required for the robust reconstruction, cf.\ 
\re{RoughNumber},
\begin{align}
P_{\rm homogeneous} \simeq T_z \prod_{k}^{L - 1}N_k
\, .
\end{align}

Comparing $P_{\rm homogeneous}$ with $P_{\rm balance}$, one immediately observes that while the number of sample points for the homogenous 
reconstruction depends on the cumulative power of the function in question, i.e., its proportionality to $T_z$, the one for the balancing method is 
controlled by the individual powers of each variable. In other words, if the function possesses a very high total power while the individual exponents 
of variables building it up are small, the balancing method will be far more effective compared to its homogenous counterpart as can be easily seen 
from Table \ref{CompTablePolynomials}.

\begin{table}[t]
\begin{center}
\begin{tabular} { | c | c | c | }
\hline
rational function & balanced & homogeneous \\
\hline
\hline
$\frac{x_1}{x_1 + x_2 + x_3}$ & 64 & 36 \\[2mm]
$x_1 x_2^2 x_3^3 + x_1 x_2 x_3 + x_3 + 10$& 67 & 156 \\[2mm]
$\frac{x_1^3 5 x_1^2 x_2^2 + x_1 x_3 + x_3 + 1}{x_1 + x_2 + x_3^2 + 1}$ & 111 & 144 \\
\hline
\end{tabular}
\end{center}
\caption{\label{CompTablePolynomials} Comparison between the numbers of sample points required for balanced and homogeneous reconstructions.}
\end{table}

Let us provide now asymptotic estimates for the number of black box probes for both methods as the number of variables tends to infinity. Introducing 
the maximal exponent $p_j$ of each $x_j$ variable in the rational function $F (\bit{x})$ as
\begin{align}
p_j =  \mbox{max} \left\{ \mbox{deg}_{x_j} [\mbox{Numerator} [F(\bit{x})]] , \mbox{deg}_{x_j} [\mbox{Denominator} [F(\bit{x})]] \right\}
\, , 
\end{align}
we will use $p_m =  \mbox{max}_j \left\{ p_j\right\}$ as their upper limit estimate. Then, according to Eq.\ \re{MinPsample}
\begin{align}
P_{\rm balance} \sim \mathcal{O} (p_1\dots p_L) \leq O(p_m^L)
\, .
\end{align}
On the other hand, the homogeneous reconstruction requires $T_z \sim \mathcal{O} (p_1 + \dots + p_L) \leq O (L p_m)$ samples on the first step,
with the other $L-1$ variables requiring $\mathcal{O} (p_m)$ probes. Cumulatively, this gives
\begin{align}
P_{\rm homogeneous} \leq O(L p_m^L)
\, .
\end{align}
Thus, the balancing method is advantageous to the homogeneous one since, in spite of a more complex organization of the algorithm, it requires less 
sample points for a robust multivariate reconstruction, especially with the growth of $L$. Its obvious disadvantage is the requirement for establishing a 
proper reconstruction order, which can however be easily achieved by means of preparatory estimates for each of the variable involved. And these are 
not time consuming.

\section{Code {\tt Reconstruction.m} and integration with FIRE}
\label{ReconFIRESection}

The algorithm introduced in Sect. \ref{BalancedReconSection} was implemented as a Mathematica code {\tt Reconstruction.m} and integrated within
the FIRE6 environment \cite{Smirnov:2019qkx} for IBP reduction of Feynman integrals. The code is attached with this submission and can be simply 
copied into the already existing {\tt fire/FIRE6/mm/} folder of FIRE6 installation. Alternatively, it is freely distributed via the repository
\begin{verbatim}
https://bitbucket.org/feynmanIntegrals/fire/src/master/FIRE6/mm/Reconstruction.m
\end{verbatim}
The main component of FIRE6 used as a input for the code is its modular arithmetic output obtained with its {\tt FIRE6p} binary to generate IBP tables  
\verb|filename_x1_.._xL_p.tables|. The file names implies that one chooses 
fixed numerical values for all variables, i.e., space-time dimension and Lorentz invariants, in the field of integer numbers modulo {\tt p} with the value of 
{\tt p} being the index of a set of hard-coded primes close to $2^{64}$. It is chosen with the {\tt \#prime} option in FIRE. The main reason to work with 
modular rather than integer arithmetic directly, is that the former is easier compared to the latter since there are only finitely many elements to deal 
with, as we explained at length in the Introduction, so that to find a solution to a given problem one could try every possibility. 

The first order of business is to perform the inverse transformation from the field of primes to rationals since sample information over distinct fields 
can be combined together with the help of the Chinese remainder algorithm \cite{Gathen2013}. It is accomplished with the command
\begin{verbatim}
RationalReconstructTables["filename_x1_..._xL_p.tables",prime_max]
\end{verbatim}
where the syntax is self-explanatory and \verb|prime_max| stands for the maximal value of {\tt p}'s used (starting from 1). The output are the tables
\verb|filename_0.tables|.

Next, the first variable {\tt x1} is reconstructed with the Thiele method using the command
\begin{verbatim}
ThieleReconstructTables["filename_x1_..._xL_0.tables", 
                        x1->Range[x1_min, x1_max]]
\end{verbatim}
from its range \verb|Range[x1_min, x1_max]|.

Analytic dependence on the remaining variables, say {\tt x2}, is found by means of the balanced Newton command
\begin{verbatim}
BalancedNewtonReconstructTables["filename_x1_x2_..._xL_0.tables", 
                                x2->Range[x2_min, x2_max],x1->x1_0]]
\end{verbatim}
for a fixed value \verb|x1_0| of the done variable, which was used to prepare the balancing tables, and {\tt x2} reconstructed from its
values in the range \verb|Range[x2_min, x2_max]|. The process is then repeated for the other $(L-2)$ {\tt x}'s.

To provide more input on the syntax of these commands, we will turn to an example in the next section along with a thorough discussion 
of the optimization of the reconstruction order.

\section{Example: three-variable reconstruction}
\label{ExampleSection}

Since there is no essential time-wise difference for the modular component of FIRE6 to handle multiloop Feynman integrals,
we choose to demonstrate details of the reconstruction procedure with a planar massless double box. The latter is parametrized by 
two Mandelstam variables $s$ and $t$ and the space-time dimension $d$, such that $\bit{x} = (d,s,t)$. A user-friendly Mathematica 
notebook accompanies this manuscript as an ancillary file along with all required scripts. All computations were done on a 10 core
MacBook Pro with Apple M1 Max silicone and 64 GB RAM.

\subsection{Preparation and estimates}
\label{EstimatesSection}

We begin with a preparation of the start file for the IBP reduction by running it in Mathematica:
\begin{verbatim}
Get["FIRE6.m"];
Internal={k1,k2};
External={p1,p2,p3};
Propagators={-k1^2,-(k1+p1+p2)^2,-k2^2,-(k2+p1+p2)^2,-(k1+p1)^2,-(k1-k2)^2,
             -(k2-p3)^2,-(k2+p1)^2,-(k1-p3)^2};
Replacements={p1^2->0,p2^2->0,p3^2->0,p1p2->-s/2,p1p3->-t/2,p2p3->1/2(s+t)};
PrepareIBP[];
Prepare[AutoDetectRestrictions->True,LI->True,PositiveIndices->7];
SaveStart["doublebox"];
\end{verbatim}
This creates {\tt doublebox.start}.

Next, we need to get a good estimate for the minimal number of sample points required for each of the three variables involved. We create a 
configuration file with the content
\begin{verbatim}
#compressor        none
#threads           1
#fthreads          1
#variables         d,s,t
#start
#folder            directory/
#problem           1 doublebox.start
#integrals         doublebox.m
#output            doublebox.tables
\end{verbatim}
where {\tt doublebox.m} refers to a Mathematica script file with a set of initial Feynman integrals chosen for the IBP reduction and determination of 
an initial set of MIs. Even though, we would typically not recommend a user to employ initial integrals with nonvanishing powers of invariant scalar 
products (the last two entries of {\tt Propagators}), it is not essential for our demonstration, so we create {\tt doublebox.m} which contains a single 
integral \verb|{1,{1,1,1,1,1,1,1,-1,-1}}|. Then, we run the bash script\footnote{The syntax used holds for the private version of FIRE6, soon to be made 
available. For the current public version, the syntax is {\tt FIRE6p -variables "$\$$d"-"$\$$s"-"$\$$t"-"$\$$p" -c doublebox -silent}.} (here for $d$)
\begin{verbatim}
#!/bin/bash
for d in {100..115}
do
  for p in {1..5}
  do
     FIRE6p -v "$d"_90_80_"$p" -c doublebox --quiet
  done
done
\end{verbatim}
to find a set of tables in the format \verb|doublebox_d0_90_80_p0.tables|. The rational reconstruction from the finite fields is then accomplished 
from these by executing the Mathematica command\footnote{There is no need to load {\tt Reconstruction.m} (separately from {\tt FIRE6.m}) before this 
evaluation as it is intrinsically integrated in FIRE6.}
\begin{verbatim}
For[d0=100,d0<=115,++d0,RationalReconstructTables[
           "doublebox_"<>ToString[d0]<>"_90_80_0.tables",5,Silent->False]]
\end{verbatim}
which generates the tables \verb|doublebox_d0_90_80_0.tables| with reconstructed rational coefficients as well as messages in how many steps this
was achieved. The one with the largest number, i.e., {\tt Rational reconstruction stable after 2 steps} implies that we needed three primes to 
do it. Finally, an estimate on $T_d$ is obtained with
\begin{verbatim}
ThieleReconstructTables["doublebox_d_90_80_0.tables",d->Range[100,115]]
\end{verbatim}
This creates an output file \verb|doublebox_d_90_80_0.tables| as well as a message {\tt Thiele re\-con\-struction stable after 11 steps}. The latter tells
us that an unambiguous Thiele reconstruction required $T_d = 12$ tables. Similar consideration are then performed for the other two variables and we 
conclude the following: three primes are needed for the rational reconstruction of both $s$ and $t$ and $T_{s,t} = 6$ is the minimal number of tables for 
their rational Thiele reconstruction.

\subsection{Rational reconstruction and Thiele}
\label{RationalAndThiele}

To start the actual reconstruction process, we need to create IBP tables making use of the above estimates for the minimal number of data points 
in each variable. These numbers depend on the order in which the recovery sequence is performed. Without any attempt to optimize it at this stage 
(we will dwell on it later in Sect.~\ref{OptimalOrderSection}), let us consider $d-s-t$ ordering. That is, we start with the variable $d$ and use 
$T_d = 12$ as a minimal number of samples in this variable, since it will be recovered with the Thiele method, while the remaining two will be 
reconstructed by means of the balanced Newton and these require only about half of the data points $N_{s,t} \simeq T_{s,2}/2$. Also to warrant a 
robust rational restoration from primes and thus to be on a safe side, we add\footnote{If one wants a faster performance, one could forego this increase.} 
an extra prime as well, i.e., we change the maximal value of {\tt p} from 4 to \verb|prime_max| = 4. As it is obvious from the naive estimate of 
$P_{\rm naive}$ in Eq.\ \re{RoughNumber}, $d-s-t$ ordering is one of the inefficient routes. 

After running the script
\begin{verbatim}
#!/bin/bash
for t in {80..83}
do
  for s in {90..93}
  do
    for d in {100..111}
    do
      for p in {1..4}
      do
        FIRE6p -v "$d"_"$s"_"$t"_"$p" -c doublebox --quiet
      done
    done
  done
done
\end{verbatim}
we generate a large list of tables \verb|doublebox_d0_s0_t0_p0.tables| with fixed integer values {\tt d0, s0, t0, p0} of all variables in their respective 
ranges. The rational reconstruction from primes is done with the Mathematica command
\begin{verbatim}
For[t0=80,t0<=83,++t0,For[s0=90,s0<=93,++s0,For[d0=100,d0<=111,++d0,
RationalReconstructTables["doublebox_"
<>ToString[d0]<>"_"<>ToString[s0]<>"_"<>ToString[t0] <>"_0.tables",4]]]]
\end{verbatim}
and results in \verb|doublebox_d0_s0_t0_0.tables|, with subsequent restoration of the variable $d$ via the Thiele method
\begin{verbatim}
For[t0=80,t0<=83,++t0,For[s0=90,s0<=93,++s0, 
ThieleReconstructTables["doublebox_d_"
<>ToString[s0]<>"_"<>ToString[t0]<>"_0.tables",d->Range[100,111]]]]
\end{verbatim}

\subsection{Balancing and balanced Newton}

Next, we turn to the balanced Newton reconstruction of the variable $s$. To this end, we have to first create its balancing tables. This is
done for a single value of the already recovered variable $d$ (below {\tt d0=100}), however, for the entire range $\alpha \in \{1, \dots, N_t\}$ 
of values $t_\alpha$ of the variable $t$ (the very same ones as used in the construction of the initial tables in Sect.~\ref{RationalAndThiele}) 
but a wider range $\beta \in \{1, \dots, T_s \}$ of values $s_\beta$ for the variable $s$ in order to be able to restore it by mean of the Thiele 
method (see Sect.~\ref{EstimatesSection}). Thus, we run the script
\begin{verbatim}
#!/bin/bash
for d in 100
do
  for t in {80..83}
  do
    for s in {90..95}
    do
      for p in {1..4}
      do
        FIRE6p -v "$d"_"$s"_"$t"_"$p" -c doublebox --quiet
      done
    done
  done
done
\end{verbatim}
with subsequent rational 
\begin{verbatim}
For[t0=80,t0<=83,++t0,
For[s0=90,s0<=95,++s0,RationalReconstructTables["doublebox_100_"
<>ToString[s0]<>"_"<>ToString[t0]<>"_0.tables",4]]]
\end{verbatim}
and Thiele reconstructions
\begin{verbatim}
For[t0=80,t0<=83,++t0,ThieleReconstructTables["doublebox_100_s_"
<>ToString[t0]<>"_0.tables",s->Range[90,95]]]
\end{verbatim}
The latter are now the {\sl balancing tables} for the variable $s$ that we sought for. Now calling the Mathematica command
\begin{verbatim}
For[t0=80,t0<=83,++t0,BalancedNewtonReconstructTables["doublebox_d_s_"
<>ToString[t0]<>"_0.tables",s->Range[90,93],d->100,Silent->False]]
\end{verbatim}
we completely reconstruct the $s$-dependence (in addition to the previously restored $d$-dependence).

Following the very same steps all over again but now for $t$, we recover it as well. The output is a file \verb|doublebox_d_s_t_0.tables| 
with full analytical dependence on all variables involved. In order to avoid being repetitive, we relegate our reader's curiocity to the 
accompanying Mathematica notebook for details.

\subsection{Optimization}
\label{OptimalOrderSection}

Finally, let us address the question of the most optimal choice for the variables' sequence during the restoration process: these are not created 
equal. In spite of the fact that the time for computation of individual probes is about 2.5 seconds\footnote{With the public version of FIRE6, this 
time is about 20 seconds.}, when many samples are needed, the total 
time it takes to compute the initial set of black box probes can get large since the growth is linear. We conducted a numerical experiment to verify 
that the time reduction factors are linearly correlated with the total number of probes required for a robust reconstruction. This was indeed confirmed 
and is reported in Table \ref{CompOrderTable}.

\begin{table}[t]
\begin{center}
\begin{tabular} { | c | c | c | }
\hline
order & min. number of tables & time reduction factor \\
\hline
\hline
$d-t-s$ & 907 & 1 \\[2mm]
$d-s-t$ & 878 & 0.96 \\[2mm]
$t-d-s$ & 854 & 0.91 \\[2mm]
$t-s-d$ & 778 & 0.83 \\[2mm]
$s-d-t$ & 582 & 0.67 \\[2mm]
$s-t-d$ & 573 & 0.66 \\
\hline
\end{tabular}
\end{center}
\caption{\label{CompOrderTable} Comparison of 3! choices for the order of functional reconstruction against the number of table required 
according to Eq.\ \re{MinPsample} and numerical experiments yielding corresponding time reduction factors.}
\end{table}

\section{Conclusions}

To conclude, in this exploratory paper, we introduced a new approach for robust reconstruction of rational functions of many variables from 
their modular arithmetic input. It is based on a balancing relation for recovery of a variable in question. The former is found from a small data set 
by means of the univariate Thiele method, which is then used in conjunction with the Newton reconstruction from a minimal original set of black box 
probes.

We developed a Mathematica language package, {\tt Reconstruction.m}, which is intrinsically integrated into the FIRE6 environment for algebraic
and modular arithmetic-based IBP reductions. We demonstrated its efficiency for a typical multiloop integral. We provided heuristic arguments 
for the most optimal choice of the multivariate reconstruction and confirmed them with numerical experiments.

A natural extension of the current work is to use it as a stepping stone for its C++ implementation along with addressing issues of optimization 
and parallelization for use on supercomputers.

\section*{Acknowledgments}

A.B.\ is deeply indebted to Vladimir Smirnov for his generous patience and meticulous explanations of inner workings of modern multiloop 
methods. We are grateful to Roman Lee for helpful correspondence and providing us with an updated version of LiteRed compatible with
Mathematica 13.2. The work of A.B. was supported by the U.S.\ National Science Foundation under the grant  No.\ PHY-2207138, while of  
A.S. and R.Y. by the Russian Science Foundation under the agreement No.\ 21-71-30003 (in part, for the development of a version of the algorithm 
applicable on supercomputers) and by the Ministry of Education and Science of the Russian Federation as a part of the program of the Moscow 
Center for Fundamental and Applied Mathematics under the agreement No.\ 075-15-2019-1621 (in part, for the development of the balancing 
reconstruction method suitable for more that two variables).


\end{document}